\documentstyle[floats,prd,aps]{revtex}

\begin{document}
\draft

\title{Three-dimensional initial data for the collision of two black
holes II: Quasi-circular orbits for equal-mass black holes}

\author{Gregory B. Cook}
\address{Center for Radiophysics and Space Research, Cornell University,
	Ithaca, New York\ \ 14853}

\date{\today}

\twocolumn[
\maketitle

\begin{abstract}
\widetext
The construction of initial-data sets representing binary black-hole
configurations in quasi-circular orbits is studied in the context of
the conformal-imaging formalism.  An effective-potential approach for
locating quasi-circular orbits is outlined for the general case of
two holes of arbitrary size and with arbitrary spins.  Such orbits
are explicitly determined for the case of two equal-sized nonrotating
holes, and the innermost stable quasi-circular orbit is located.
The characteristics of this innermost orbit are compared to previous
estimates for it, and the entire sequence of quasi-circular orbits is
compared to results from the post-Newtonian approximation.  Some
aspects of the numerical evolution of such data sets are explored.
\end{abstract}
]

\narrowtext
\section{Introduction}
\label{sec:intro}
The numerical simulation of binary black-hole systems begins with
the specification of appropriate initial data.  A general method
for specifying the initial data of binary black-hole systems has
been described in Cook {\it et al.}\cite{cook93} (hereafter Paper~I).
Because of the
circularizing effects of gravitational radiation damping, we expect
the orbits of most tight binary systems to have small eccentricities.
Therefore, a method is needed which can discern which data sets,
within the very large parameter space of binary black-hole
initial-data sets, correspond to binary black holes in a quasi-circular
orbit.  In this paper, I will describe such a method and explicitly
compute the initial-data parameters necessary for describing the
quasi-circular orbit of two equal-sized nonrotating black holes.
In addition, this method yields an estimate of the innermost stable
quasi-circular orbit for two equal-sized black holes.

The general framework being used for defining initial-data sets
containing black holes is known as the {\em conformal-imaging
formalism}\cite{york79,bowen79,bowenyork80,bowen82,ksy83,cook91}.
It is based on the Arnowitt-Deser-Misner
(ADM)\cite{ADM}, or 3+1, decomposition of Einstein's equations,
York's conformal and transverse-traceless decompositions of the
constraint equations, and a method of imaging applicable to tensors.
Application of this approach to the case of two black holes with
arbitrary linear and angular momenta on each hole has been explored
in Paper~I.  In that work, three independent
numerical approaches were described for constructing initial-data sets.
I will make use, in this paper, of the ``\v{C}ade\v{z}-coordinate
approach'' for solving the Hamiltonian constraint, a
three-dimensional (3D) quasi-linear elliptic partial differential
equation.

Using numerical initial-data sets, an ``effective potential'' can be
constructed which consists of the gravitational binding energy between
the holes plus the kinetic energy of the holes, with the restriction
that all physical parameters characterizing the system are held
fixed except for the separation of the holes.  If we fix the momenta
of the holes appropriately, then an initial-data set occurring at a
minimum of the effective potential will represent two black holes in
a quasi-circular orbit.  For the case of two equal-sized holes with
no intrinsic rotation, the set of physical parameters that must be
held fixed contains only the orbital angular momentum of the system.
For this simple case, we have located those initial-data sets which
represent two black holes in quasi-circular orbit at various
separations.  As expected\cite{detweiler92,kidder92}, there is
some minimum separation required before a minimum of the effective
potential exists.  This minimum separation serves as an estimate for
the point at which the secular inspiral of the two holes ends and the
final dynamical plunge to coalescence occurs.

I will begin with a description of the parameter space
of 3D initial-data sets, a brief description of how initial-data sets
are constructed, and how the physical parameters associated with a
given data set are computed.  I will continue with a description of
the effective-potential method as applied to initial data constructed
via the conformal-imaging approach.  Using this method, I explicitly
locate the sequence of quasi-circular orbits (as a function of
separation) for equal-sized holes and compare this sequence to the
results obtained via a post-Newtonian analysis.  Also, the
characteristics of the innermost stable quasi-circular orbit are
compared to the results of previous estimates for this orbit.
I conclude with some observations about the eventual
evolution of initial-data sets representing two equal-sized holes
in quasi-circular orbit.

\section{Initial data}
\label{sec:id}
A detailed description of how binary black-hole initial-data sets
are constructed within the conformal-imaging formalism is given
in Paper~I and references therein.  In brief, the
configuration is parameterized first by fixing the locations,
${\bf C}_1$ and ${\bf C}_2$, of the two holes in a Cartesian flat
conformal {\em background} space along with the radii, $a_1$ and
$a_2$, of the two holes.  If we let $a_1$ set the fundamental length
scale of the problem, then we can parameterize the physical locations
and relative sizes of the holes in the background space by two
dimensionless parameters, $\alpha$ and $\beta$, defined by
\begin{eqnarray}
\label{eqn:alpha_def}
	\alpha &\equiv& {a_1 \over a_2} \\
\label{eqn:beta_def}
	\beta &\equiv& {|{\bf C}_1 - {\bf C}_2| \over a_1} .
\end{eqnarray}

In addition to the locations and sizes of the holes, we also specify
for each hole both a linear momentum vector ${\bf P}_{1,2}$ and
an angular momentum vector ${\bf S}_{1,2}$.  The physical meaning
of these momenta is seen in the following way.  For a single isolated
hole, the {\em physical} linear and angular momenta contained in the
initial-data set, as measured at infinity, are given directly by the
parameters ${\bf P}$ and ${\bf S}$.  For two holes, the total physical
linear momentum of the system is simply the vector sum
${\bf P}_1 + {\bf P}_2$.  The total physical angular momentum is
given by the vector sum ${\bf S}_1 + {\bf S}_2 + {\bf J}$, where
${\bf J}$ is the orbital angular momentum of the system which will
be described in detail below.  With $a_1$ setting the fundamental
length scale, we find that the momenta of two black holes within
the initial-data set is completely specified by the four dimensionless
vector parameters: ${\bf P}_1/a_1$, ${\bf P}_2/a_1$, ${\bf S}_1/a_1^2$,
and ${\bf S}_2/a_1^2$.

Aside from the complications of a given numerical technique used
to solve for the initial data, we see that the specification of a
general two-hole configuration requires that fourteen dimensionless
parameters be fixed.  Having chosen these fourteen parameters, an
initial-data set is constructed by solving the momentum and
Hamiltonian constraints of general relativity as outlined in
Paper~I.  The particular numerical technique used to
solve the Hamiltonian constraint in this work is the
``\v{C}ade\v{z}-coordinate approach'' described in detail in
Sec.~IIIA of that paper and references therein.

Once the initial-data set has been computed, various physical
parameters characterizing the data set can be computed.  In
particular, the following quantities are calculated: the ADM
energy of the system $E$, the proper surface areas of the
marginally outer-trapped surfaces defining each individual hole
$A_{1,2}$, the dipole moment of the energy distribution ${\bf d}$,
and the shortest proper separation between the two marginally
outer-trapped surfaces $\ell$.  Definitions for the ADM energy
and the dipole moment can be found in Eqns.~(24) and (25) of
Ref.~\cite{cook91}.  Note that it is actually the dimensionless
ratios $E/a_1$, $A_{1,2}/a_1^2$, ${\bf d}/a_1^2$, and $\ell/a_1$ which
are computed.  These quantities, together with the initial-data
parameters, allow us to compute the orbital angular momentum ${\bf J}$
of the system.  Generalizing the calculation found in
Ref.~\cite{Frontiers:York}, it is straightforward to show that
the orbital angular momentum for a configuration of two black
holes is given by
\begin{equation}
\label{eqn:J_gen}
	{\bf J} = ({\bf C}_1 - {\bf O})\times{\bf P}_1 +
		({\bf C}_2 - {\bf O})\times{\bf P}_2 ,
\end{equation}
where ${\bf O}$ is the point in the background space about which
the angular momentum is defined.  The only unique choice for
${\bf O}$ is the center of energy of the system which,
if chosen as the origin of coordinates, results in
a vanishing dipole moment.  Using the definition of the ADM energy
and the dipole moment, it follows that the center of energy is at
${\bf O} = {\bf d}/E$.  Therefore, the general definition of the
orbital angular momentum is
\begin{equation}
\label{eqn:J}
	{\bf J} \equiv \left({\bf C}_1 - {{\bf d}\over E}\right)
						\times{\bf P}_1 +
			\left({\bf C}_2 - {{\bf d}\over E}\right)
						\times{\bf P}_2 .
\end{equation}

\section{The effective-potential method}
\label{sec:effect_pot}
The definitions of the ADM energy, total linear and angular
momenta at infinity, the dipole moment, the proper separation
of the holes, and the areas of the two marginally outer-trapped
surfaces are rigorously defined physical quantities.
To define the effective potential of a configuration, it is necessary
to use the concepts of the masses and spins of the individual black
holes and to have a measure of the effective binding energy between
the two holes.  However, none of these quantities are rigorously
defined in general relativity for a strong-field nonstationary
configuration.

In the limit of large separations and small linear momenta and spin
on the holes, the following definitions hold.  The mass of each hole
can be defined via the Christodoulou formula \cite{christ70}
\begin{equation}
\label{eqn:m_christ}
	M^2 = M^2_{\rm ir} + {S^2 \over 4M^2_{\rm ir}} ,
\end{equation}
with the irreducible mass $M_{\rm ir} \approx \sqrt{A/16\pi}$ and
$S$ being the magnitude of the spin on the hole.  As shown in the
Appendix, we can approximate $S$ for each hole by the magnitude of
its respective spin parameter $|{\bf S}_{1,2}|$.  Though it seems
possible that the linear momentum on one hole could {\em induce}
a spin on the other hole, this is, in fact, not the case.
Thus, the quantity ${\bf J}$, which we identified above as the orbital
angular momentum, is not contaminated by an induced spin on
the holes.  We are therefore justified in defining the orbital
angular momentum of the system by equation (\ref{eqn:J}), and we
define the spins of the individual holes via their respective
spin parameters.  Finally, the effective binding energy $E_{\rm b}$
between the holes is defined as
\begin{equation}
\label{eqn:Eb}
		E_{\rm b} \equiv E - M_1 - M_2 ,
\end{equation}
where $M_{1,2}$ are the masses of the two holes as defined above.
Note that the effective binding energy contains both the gravitational
binding energy between the two holes and their orbital kinetic
energies, but not the rotational kinetic energy of the individual holes.

These definitions are rigorous only in the limit of infinite
separation and zero momenta on either hole.  The limit of zero momenta
on the holes is required because initial-data sets containing a single
black hole constructed via the conformal-imaging approach necessarily
contain some spurious gravitational-wave energy \cite{cook90}.  The
same is true of multihole initial-data sets, however, Cook and
Abrahams\cite{cookabrahams92} have shown that the spurious radiation
content for the case of two holes is quite small so long as the holes
are modestly separated and the momenta are not excessively large.
We will find that these constraints are satisfied for the majority
of configurations of physical interest.

Henceforth, we will take the masses of the holes, their spins,
mutual binding energy, and orbital angular momentum to be
{\em defined} as given above.

We turn now to the definition of an effective potential useful for
determining the location of quasi-circular orbits.  In general, such
an effective potential should be a function of the separation and
sizes of the holes, the orbital angular momentum of the system, the
spins of the holes, and the gravitational radiation content of the
system.  Within the conformal-imaging approach, one has no freedom
in specifying the radiation content of the system.  This is fixed
by the demands that the spatial 3-metric be conformally flat and that
all fields satisfy an isometry condition (cf.\ Ref.~\cite{cook91}).
With this restriction, we see that the effective potential should
be a function of {\em nine} physical parameters but a general
two-hole initial-data set depends on fourteen initial-data parameters.

To reduce the size of this parameter space, we first demand that the
configuration be in a center of momentum frame.  This restriction
requires
\begin{equation}
\label{eqn:cmf}
	{\bf P}_1 + {\bf P}_2 = 0.
\end{equation}
A configuration representing a quasi-circular orbit should satisfy
\begin{equation}
\label{eqn:qsodef}
	{\bf P}_{1,2}\cdot({\bf C}_2 - {\bf C}_1) = 0.
\end{equation}
Together, equations (\ref{eqn:cmf}) and (\ref{eqn:qsodef}) reduce the
fourteen-dimensional initial-data parameter space to nine dimensions.
These parameters are $\alpha$, $\beta$, the magnitude of the linear
momentum on either hole $P/a_1$, ${\bf S}_1/a_1^2$, and
${\bf S}_2/a_1^2$.

The respective dimensionless physical parameters of the effective
potential are $X \equiv M_1/M_2$, $\ell/m$, $J/\mu{m}$,
${\bf S}_1/M_1^2$, and ${\bf S}_2/M_2^2$ where $M_{1,2}$ are the
masses of the individual holes, $m \equiv M_1 + M_2$ is the total
mass, $\mu \equiv M_1M_2/m$ is the reduced mass, and $J$ is the
magnitude of the orbital angular momentum of the system.  Finally,
the dimensionless effective potential is given by the binding
energy as $E_{\rm b}/\mu$.

Finding quasi-circular orbits is now a conceptually easy task.  We
compute $E_{\rm b}/\mu$ as a function of $\ell/m$ while holding the
remaining eight {\it physical} parameters constant.  A minimum in
$E_{\rm b}/\mu$ then corresponds to a ``stable'' quasi-circular orbit.
In addition to locating the quasi-circular orbits, we can also
estimate the orbital angular velocity $\Omega$ of the system as
measured at infinity.  This is given by taking the derivative of
the binding energy with respect to the orbital angular momentum
while holding all other parameters fixed.  In dimensionless form
then one obtains
\begin{equation}
\label{eqn:Omega}
	m\Omega = {\partial E_{\rm b}/\mu\over\partial J/\mu{m}} .
\end{equation}

Though conceptually straightforward, the computational task of
locating quasi-circular orbits is difficult.  The main difficulty
arises from the fact that the physical parameters that must be
held fixed are not independently correlated with their respective
initial-data parameters.  That is, holding eight of the nine
initial-data parameters fixed while varying $\beta$ will not result
in an effective-potential curve.  We see then that, in general, the
problem of determining {\em one} quasi-circular orbit is quite
involved.  It requires finding the roots of eight functions, each
of which depends on eight parameters, at each value of the separation
at which the effective potential is evaluated.

Fortunately, the size of this parameter space can be cut in half.
The only definition we have available for the direction of the
spins of the holes (see the Appendix) is the direction of their
respective initial-data spin parameters.  The
directions of the spins are then fixed relative to the separation
of the holes and the plane of the orbit, which are defined by
${\bf P}_{1,2}$ and ${\bf C}_2 - {\bf C}_1$.  This direction is
independent of the numerical solution of the Hamiltonian constraint,
so only the magnitude of the two initial-data spin-vector
parameters needs to be varied to hold the physical
parameters fixed.  As a result, for a fixed value of $\beta$
the following four equations must be satisfied:
\begin{mathletters}
\label{eqn:roots}
\begin{eqnarray}
\label{eqn:X_root}
	X_{(\alpha,P/a_1,S_1/a_1^2,S_2/a_1^2;\beta)} &=& X_0 \\
\label{eqn:J_root}
	\left[{J\over\mu{m}}\right]_
		{(\alpha,P/a_1,S_1/a_1^2,S_2/a_1^2;\beta)} &=&
			\left.{J\over\mu{m}}\right|_0 \\
\label{eqn:S1_root}
	\left[{S_1\over M_1^2}\right]_
		{(\alpha,P/a_1,S_1/a_1^2,S_2/a_1^2;\beta)} &=&
			 \left.{S_1\over M_1^2}\right|_0 \\
\label{eqn:S2_root}
	\left[{S_2\over M_2^2}\right]_
		{(\alpha,P/a_1,S_1/a_1^2,S_2/a_1^2;\beta)} &=&
			 \left.{S_2\over M_2^2}\right|_0 .
\end{eqnarray}
\end{mathletters}
When these four equations are satisfied, equation (\ref{eqn:Eb})
yields a value of the effective potential $E_{\rm b}/\mu$ at
some value of the physical separation $\ell/m$.  Changing the value
of $\beta$ and resolving
Eqns.~(\ref{eqn:X_root})--(\ref{eqn:S2_root})
results in another value of the effective potential at a different
separation.

\section{Equal-mass nonrotating holes}
\label{sec:equal_holes}
The simplest application of the effective-potential method is to
the case of two equal-sized black holes with no intrinsic spin.  In
this problem, $S_1/a_1^2 = S_2/a_1^2 = 0$ and, because of the symmetry
between the holes, we know that $\alpha = 1 \to X = 1$.  Therefore,
solving for the effective potential requires solving only
Eqn.~(\ref{eqn:J_root}) as a function of $P/a_1$ alone for a given
$\beta$.

The method used to solve equation (\ref{eqn:J_root}) is the following.
For a given value of $\beta$, the initial-value equations are solved
at a sufficiently large number of values of $P/a_1$ in order to
encompass all of the values of $J/\mu{m}$ at which we want to evaluate
the effective potential.  Using interpolation, we can estimate new
values of $P/a_1$ that will yield solutions near the desired values
of $J/\mu{m}$; the procedure is repeated until any errors
introduced by interpolation are sufficiently small.  The most difficult
part of the problem is to solve the initial-value equations with
sufficient accuracy.  Typically, both the ADM mass $E/a_1$ and the
areas of the marginally outer-trapped surfaces $A_{1,2}/a_1^2$ need
to be determined to a relative error of $\sim 10^{-5}$.

Currently, the only numerical method capable of solving the
initial-value equations to this accuracy is the multigrid-based
``\v{C}ade\v{z}-coordinate approach'' described in detail in
section IIIA of Paper~I.  Such high accuracy can be
obtained through the use of Richardson extrapolation.  As described
in Paper~I, the \v{C}ade\v{z}-coordinate approach used to solve the
Hamiltonian constraint results in a numerical solution for the
conformal factor $\psi^{\rm num}$, which has an asymptotic
($h \to 0$) expansion given by
\begin{equation}
\label{eqn:psi_exp}
	\psi^{\rm num} = \psi + h^2 e_2^\psi + h^4 e_4^\psi + \cdots ,
\end{equation}
where $\psi$ is the analytic solution of the Hamiltonian constraint,
$h$ is the basic scale of discretization, and
$e_2^\psi,e_4^\psi,\ldots$ are $h$-independent functions.
In addition, the numerical integrals for $E/a_1$, ${\bf d}/a_1^2$,
$A_{1,2}/a_1^2$, and $\ell/a_1$ have all been constructed to
yield analogous error expansions that depend strictly on powers
of $h^2$.

One final source of error which must be examined comes from the
necessity of imposing an approximate outer-boundary condition
(cf. Ref.~\cite{cook91}).  In order to minimize the effects of
this approximation, the outer boundary has been placed at a
distance of at least $2000a_1$ from the holes.

\begin{figure}
\special{hscale=0.45 vscale = 0.45 hoffset = -0.1 voffset = -3.35
psfile=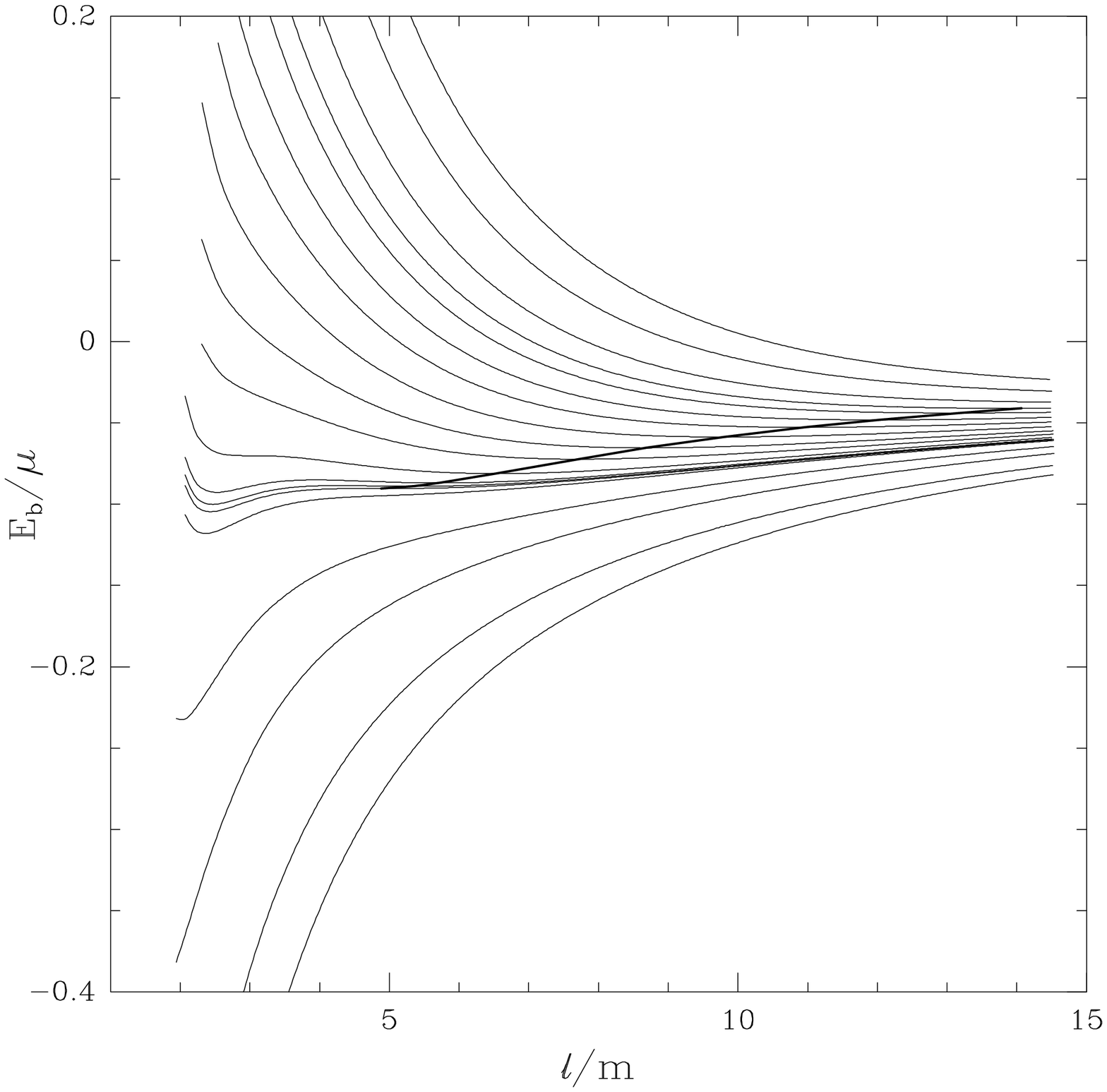}
\vspace{3.4in}
\caption{The effective potential $E_{\rm b}/\mu$ as a function of
separation $\ell/m$ for the following values of the orbital angular
momentum $J/\mu{m}$: 1.5, 2, 2.5, 2.75, 2.95, 2.976, 2.985, 3, 3.05,
3.15, 3.25, 3.37, 3.5, 3.62, 3.75, 3.85, 4, 4.25, and 4.5.  These
values of $J/\mu{m}$ label, respectively, curves from the bottom of
the figure to the top.  Also, plotted as a bold line is the sequence
of quasi-circular orbits which cross the effective-potential curves
at local minima.}
\label{fig:Eb_v_L:all}
\end{figure}

\begin{figure}
\special{hscale=0.45 vscale = 0.45 hoffset = -0.1 voffset = -3.35
psfile=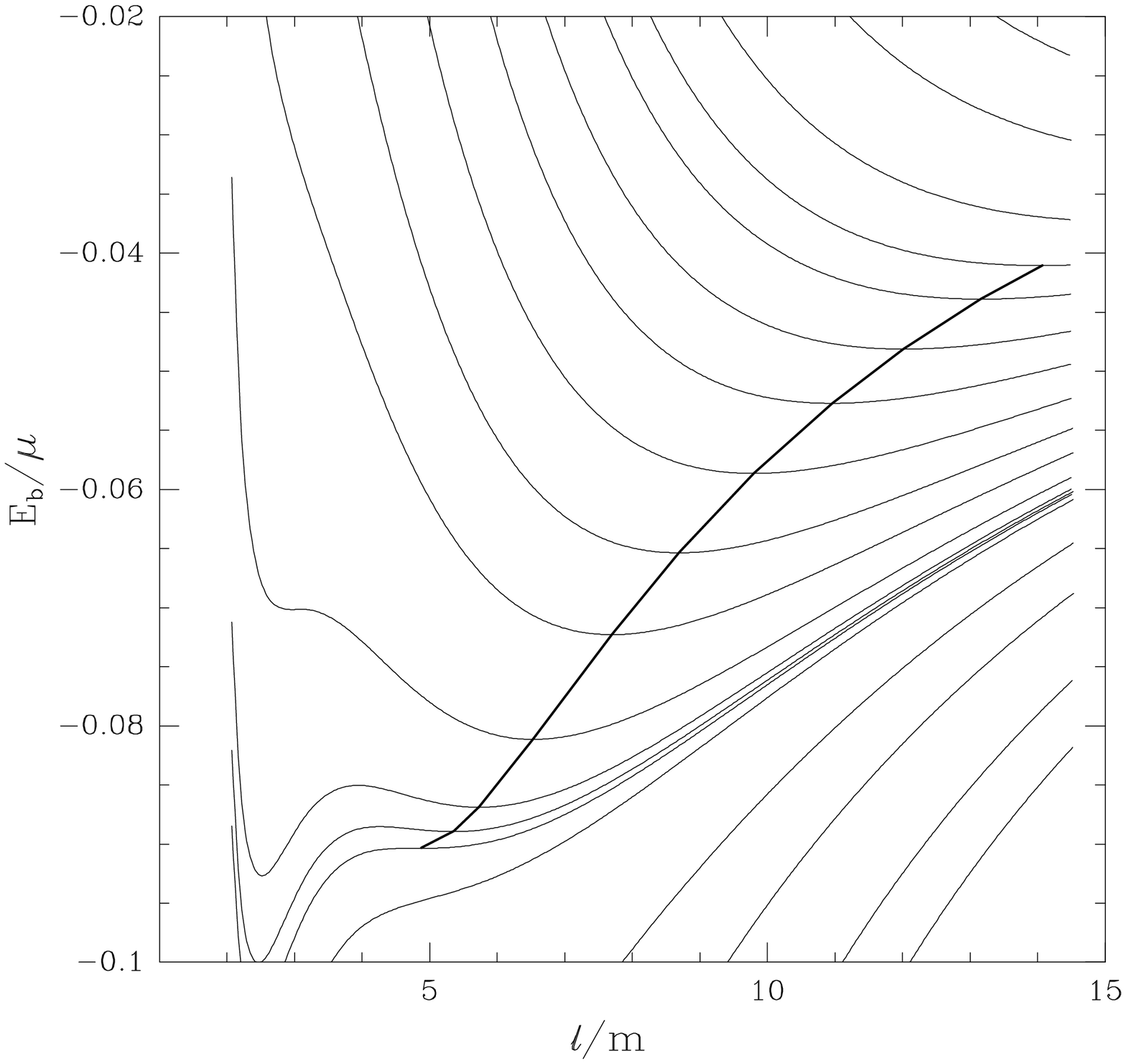}
\vspace{3.3in}
\caption{An enlargement of the section of
Fig.~\protect\ref{fig:Eb_v_L:all}
which contains the sequence of quasi-circular orbits.  This sequence
begins at the innermost stable quasi-circular orbit near $\ell/m = 5$
on the $J/\mu{m} = 2.976$ curve and extends in the direction of
larger separation.  In the figure, the sequence terminates at the
minimum of the $J/\mu{m} = 3.85$ effective-potential curve, although
it should continue.}
\label{fig:Eb_v_L:qso}
\end{figure}

Figure~\ref{fig:Eb_v_L:all} displays a set of effective-potential
curves for a wide range of values for $J/\mu{m}$.
The displayed curves are interpolated results derived from
3000 Richardson-extrapolated data points, each resulting from
the extrapolation of three separate solutions of the initial-value
equations at resolutions similar to those described in
Paper~I.  All solutions were generated on an
IBM-SP1 parallel computer and required a total computational time
in excess of 3000 CPU hours.
The value of $J/\mu{m}$ is held fixed along each of the
thin curves which plots the effective potential $E_{\rm b}/\mu$
as a function of the proper separation of the holes $\ell/m$.
The bold curve crossing several of the effective-potential
curves represents a sequence of quasi-circular orbits.  This can
be seen more clearly in Fig.~\ref{fig:Eb_v_L:qso} where the
region containing minima in the effective-potential curves
is shown in expanded form.
The bold line representing the
sequence of quasi-circular orbits begins at the right at an
$\ell/m \sim 14$.  This line should, of course, extend to larger
values of $\ell/m$, but data has not been computed in this regime.
Following the sequence of quasi-circular orbits to smaller values
of $\ell/m$, we find that the minimum in the effective potential
vanishes.  At this point, the sequence of quasi-circular orbits
terminates at an innermost stable orbit.

For $\ell/m \lesssim 4$ we notice that some of the
effective-potential curves pass through a local maximum and a second
minimum.  These additional minima should not be interpreted as
a new sequence of stable quasi-circular orbits.  Rather, this
behavior indicates that the approximations outlined in
Sec.~\ref{sec:effect_pot}, especially the identification of
equation (\ref{eqn:Eb}) as a measure of the binding energy between
the holes, are breaking down.  This assertion is justified on
the following grounds.  Anninos {\it et al.}~\cite{anninos_etal94}
have shown that for time-symmetric initial data constructed via
the conformal-imaging approach an {\em event horizon} encompasses
both holes on the initial-data slice when their separation parameter
$\mu$ (not to be confused with the reduced mass) is less than about
1.8.  This corresponds to a separation
parameter of $\beta \sim 6.2$ and to a proper separation of
$\ell/m \sim 3$.  For the case of orbiting holes, we expect
the effects of angular momentum to delay the onset of
formation of a common event horizon based on both physical
intuition and the axisymmetric results of Cook and
Abrahams\cite{cookabrahams92}.  Thus,
we should expect a common event horizon for $\ell/m < 3$ and,
further, that the effective-potential method being used should
not be trusted for $\ell/m < 4$.

The characteristics of initial-data sets representing quasi-circular
orbits of two equal-sized, nonrotating black holes are given in
Table~\ref{tab:qso_params}.
\begin{table}
\caption{Physical and initial-data parameters characterizing certain
configurations along the sequence of stable quasi-circular orbits.
The data sets represented in this table have been constructed using
the {\it minus} isometry condition of the conformal-imaging approach.
}
\label{tab:qso_params}
\begin{tabular}{dccllc}
$\ell/{m}$&$E_{\rm b}/\mu$&$J/\mu{m}$&$\,m\Omega$&$\,P/a_1$&$\beta$
\hbox{\,\,} \\
\tableline
4.880 & $-$0.09030 & 2.976 & 0.172  & 1.685  & 11.82 \hbox{\,\,}  \\
5.365 & $-$0.08890 & 2.985 & 0.145  & 1.392  & 13.28 \hbox{\,\,}  \\
5.735 & $-$0.08684 & 3.000 & 0.130  & 1.230  & 14.43 \hbox{\,\,}  \\
6.535 & $-$0.08112 & 3.050 & 0.104  & 0.9868 & 16.99 \hbox{\,\,}  \\
7.700 & $-$0.07226 & 3.150 & 0.0774 & 0.7752 & 20.84 \hbox{\,\,}  \\
8.695 & $-$0.06534 & 3.250 & 0.0622 & 0.6613 & 24.21 \hbox{\,\,}  \\
9.800 & $-$0.05862 & 3.370 & 0.0504 & 0.5734 & 28.04 \hbox{\,\,}  \\
10.96 & $-$0.05270 & 3.500 & 0.0414 & 0.5066 & 32.15 \hbox{\,\,}  \\
12.02 & $-$0.04810 & 3.620 & 0.0352 & 0.4609 & 35.93 \hbox{\,\,}  \\
13.16 & $-$0.04388 & 3.750 & 0.0300 & 0.4218 & 40.06 \hbox{\,\,}  \\
14.07 & $-$0.04104 & 3.850 & 0.0270 & 0.3960 & 43.38 \hbox{\,\,}  \\
\end{tabular}
\end{table}
In addition to the physical parameters
characterizing the system ($\ell/m$, $E_{\rm b}/\mu$, $J/\mu{m}$,
and $m\Omega$), the table contains the initial-data parameters
$P/a_1$ and $\beta$ required to reproduce these particular data sets.
Note that all initial-data sets described in this paper satisfy the
{\it minus} isometry condition of the conformal-imaging approach.
Values in this table should be considered accurate to better than 1\%
with the exception of $m\Omega$, which should be considered accurate
to a few percent.

In order to gauge the accuracy with which we have located the
innermost stable quasi-circular orbit, note that in the limit
of a test mass orbiting a Schwarzschild black hole, the proper
separation between the event horizon and the test mass is
found to be
\begin{equation}
\label{eqn:testmass}
	{\ell\over m} = 2\ln\left({1 + \sqrt{2/3}\over
			1 - \sqrt{2/3}}\right) \approx 4.58 .
\end{equation}
This should be compared with a value of $\ell/m = 4.88$ obtained
from Table~\ref{tab:qso_params}.  The ratio of these two values is
$0.94$.  Kidder {\it et al.}\cite{kidder92} obtain an analogous
ratio of $0.96$, where separation is measured in terms of harmonic
or deDonder coordinates.  Kidder {\it et al.} obtain this result
via a critical point analysis of the equations of motion through
(post)${}^2$-Newtonian order.  After altering the equations of
motion significantly to reproduce exactly the test mass
limit, Kidder {\it et al.} find that the ratio has changed to
$0.83$ and that the innermost stable
circular orbit is characterized by the following physical
parameters: $E_{\rm b}/\mu \sim -0.0378$, $J/\mu{m} \sim 3.83$,
and $m\Omega \sim 0.0605$.  A comparison of these values with
Table~\ref{tab:qso_params} shows that Kidder {\it et al.} are
finding an innermost stable circular orbit in which the holes
are much farther apart than we find with the effective-potential
method.  In contrast to this, Blackburn and
Detweiler~\cite{detweiler92} have used a variational principle
together with the assumption of a periodic solution to Einstein's
equations to obtain an estimate for the innermost orbit for two
equal-sized holes.  Using a single trial geometry, which they call
``rather unsophisticated'', they obtain an innermost orbit
characterized by $E_{\rm b}/\mu \sim -0.65$, $J/\mu{m} \sim 0.85$,
and $m\Omega \sim 2$.  Such an orbit is much more tightly bound
than seems possible from the effective-potential method.  Blackburn
and Detweiler point out that the assumptions of their variational
principle have been violated by the time this innermost circular
orbit is reached.  However, they describe a less tightly bound
circular orbit that should not be in violation of the underlying
assumptions of their approach.  This orbit has a binding energy of
$E_{\rm b}/\mu \sim -0.28$, which is still three times larger than
the binding energy obtained in this paper for the innermost stable
quasi-circular orbit.

\section{The post-Newtonian limit}
\label{sec:post_newton}
Comparison with previous estimates for the innermost stable
quasi-circular orbit of two equal-mass nonrotating black holes is
far from yielding a consensus as to its proper value.  However, we
can gain some insight into the reliability of the results derived
in this paper by comparing the sequence of quasi-circular orbits
against the post-Newtonian description of circular orbits.  Based on
the (post)${}^2$-Newtonian results of Kidder
{\it et al.}\cite{kidder93} for the binding energy, angular momentum,
and equations of motion for a binary system with a circular orbit,
it is straightforward to show that the binding energy and  angular
momentum of two compact objects in a circular orbit must satisfy
\begin{eqnarray}
\label{eqn:pn_EvJ}
	{E_{\rm b}\over\mu} &=& -{1\over2}\left({\mu{m}\over J}\right)^2
		\Bigg[1 + {1\over4}(9 + \eta)
				\left({\mu{m}\over J}\right)^2
	\nonumber\\ & &\makebox[0.04in]{} +
		\left({81\over8}-{7\over8}\eta+{1\over8}\eta^2\right)
				\left({\mu{m}\over J}\right)^4 +
		\cdots \Bigg]  \\
\label{eqn:pn_EvO}
	{E_{\rm b}\over\mu} &=& -{1\over2}\left(m\Omega\right)^{2/3}
		\Bigg[1 - {1\over12}(9 + \eta)
				\left(m\Omega\right)^{2/3}
	\nonumber\\ & &\makebox[0.04in]{} -
		\left({27\over8}-{19\over8}\eta+{1\over24}\eta^2\right)
				\left(m\Omega\right)^{4/3} +
		\cdots \Bigg]  \\
\label{eqn:pn_JvO}
	\left({J\over\mu{m}}\right)^{\!\!2}\!\! &=&
		\left(m\Omega\right)^{-2/3}
		\Bigg[1 + {1\over3}(9 + \eta)
				\left(m\Omega\right)^{2/3}
	\nonumber\\ & &\makebox[0.04in]{} +
		\left(9-{17\over4}\eta+{1\over9}\eta^2\right)
				\left(m\Omega\right)^{4/3} +
		\cdots \Bigg] ,
\end{eqnarray}
where $\eta\equiv\mu/m$.  The three terms inside the square brackets
represent the Newtonian results for circular orbits along with the
first and second post-Newtonian corrections.

\begin{figure}
\special{hscale=0.45 vscale = 0.45 hoffset = -0.1 voffset = -3.2
psfile=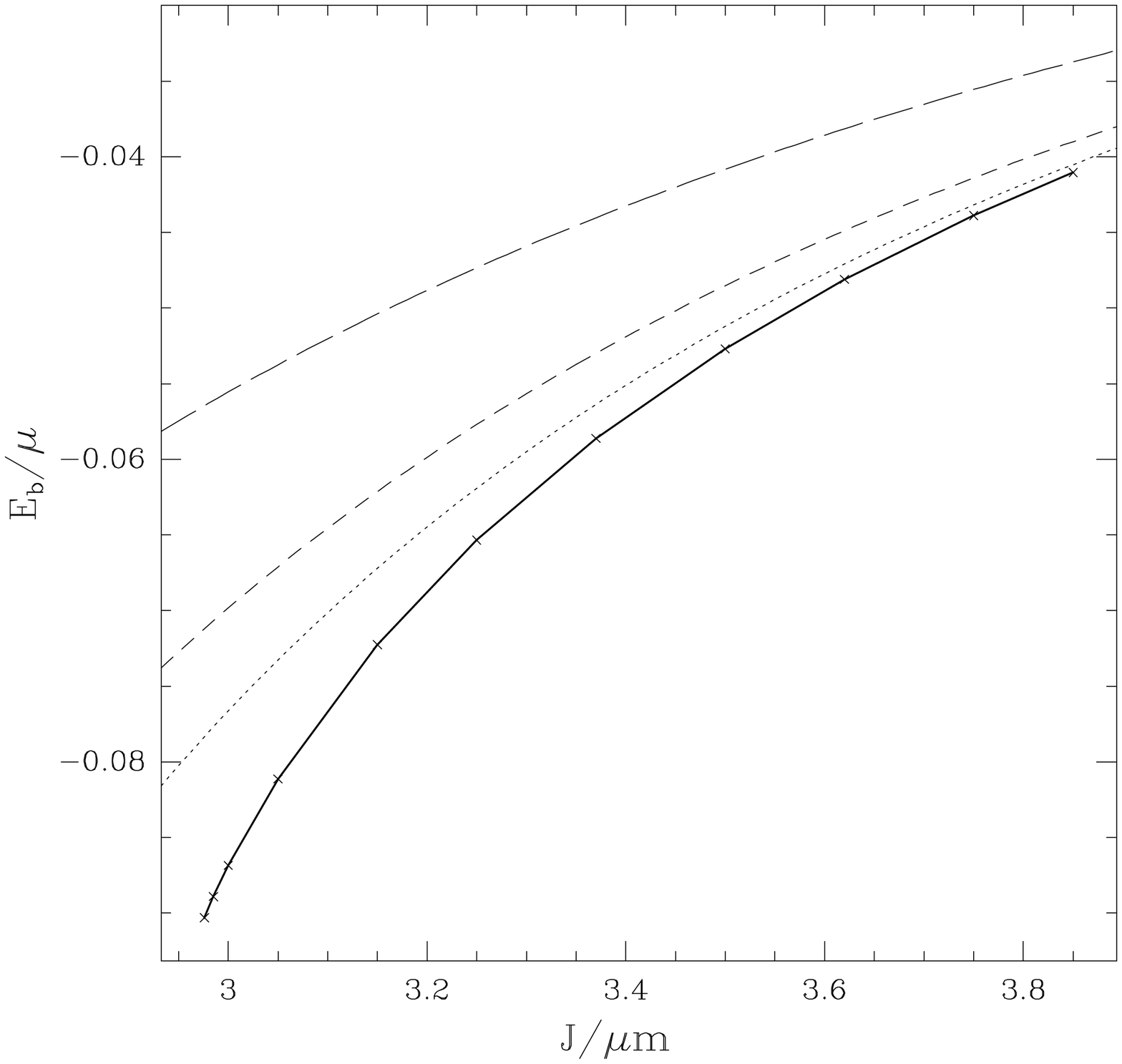}
\vspace{3.2in}
\caption{The effective potential $E_{\rm b}/\mu$ as a funcion of
the orbital angular momentum $J/\mu{m}$ for quasi-circular orbits.
The solid line corresponds to the sequence of quasi-circular orbits
computed in this paper.  The long dashed line is the result obtained
from Newtonian theory.  The short dashed line is the result based on
(post)${}^1$-Newtonian theory, and the dotted line is the result
based on (post)${}^2$-Newtonian theory.}
\label{fig:Eb_v_J}
\end{figure}

\begin{figure}
\special{hscale=0.45 vscale = 0.45 hoffset = -0.1 voffset = -3.15
psfile=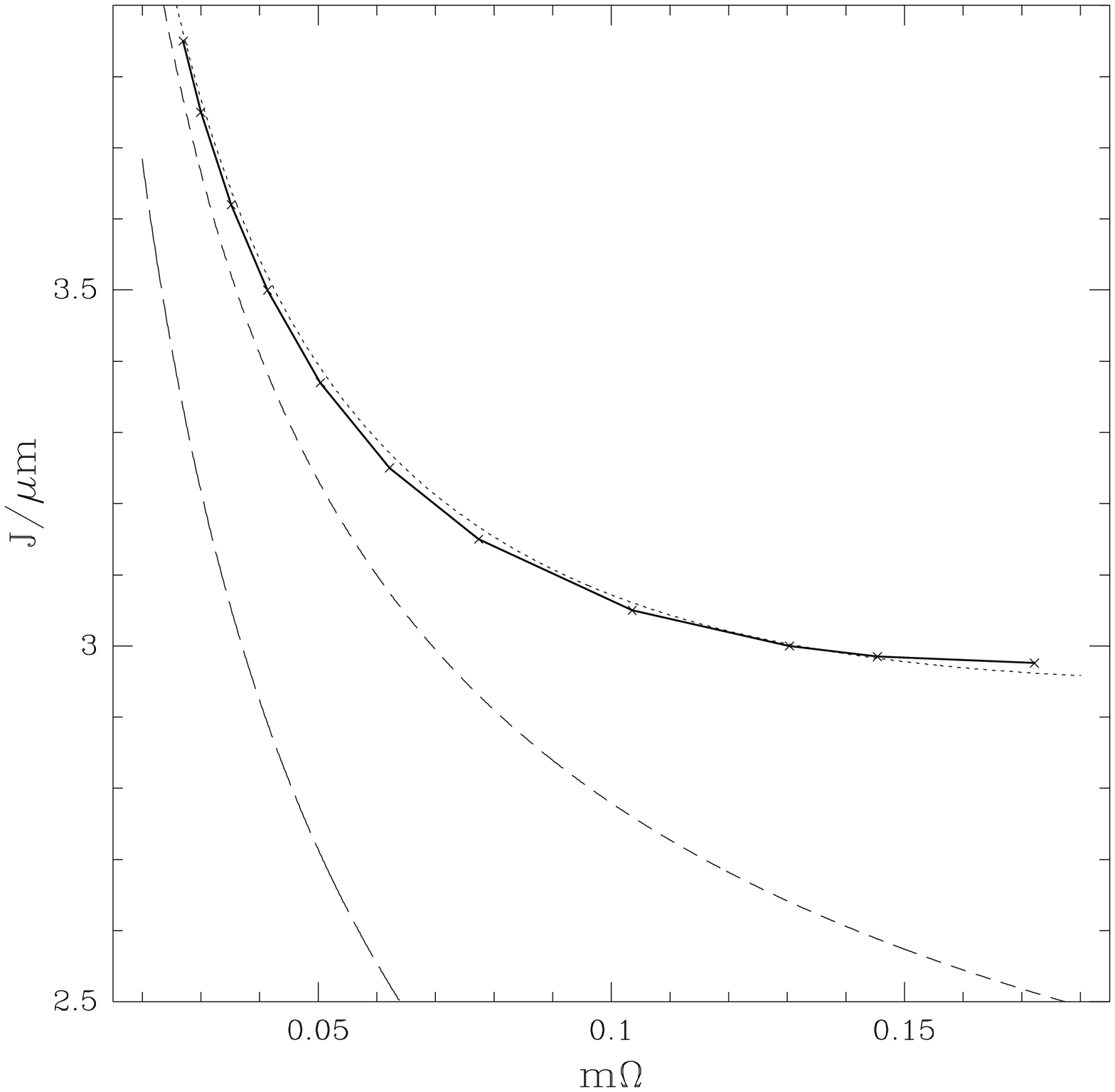}
\vspace{3.2in}
\caption{The orbital angular momentum $J/\mu{m}$ as a function of
the orbital angular velocity $m\Omega$.  Lines are as indicated in
Fig.~\protect\ref{fig:Eb_v_J}.}
\label{fig:J_v_Omg}
\end{figure}

\begin{figure}
\special{hscale=0.45 vscale = 0.45 hoffset = -0.1 voffset = -3.4
psfile=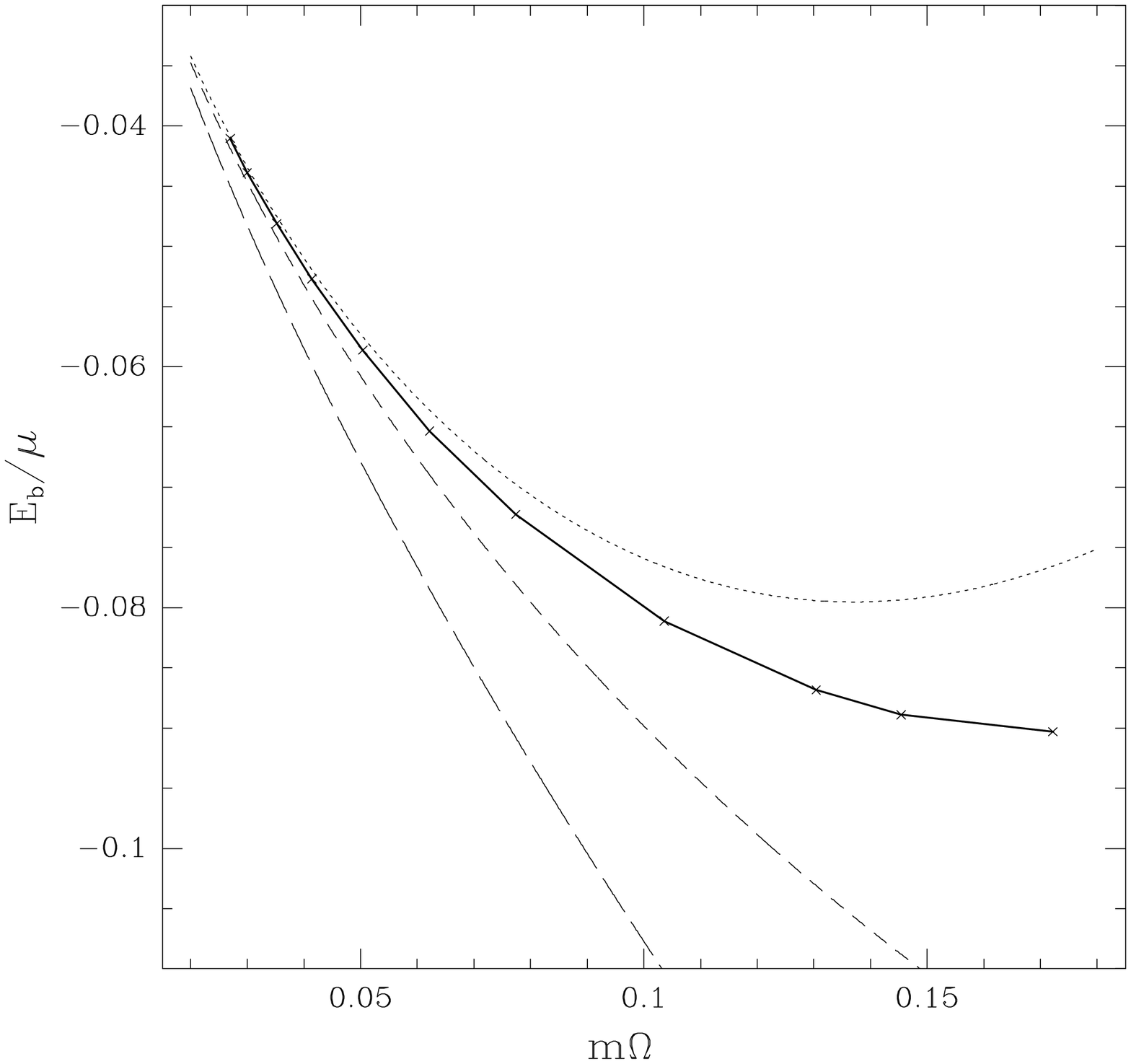}
\vspace{3.3in}
\caption{The effective potential $E_{\rm b}/\mu$ as a function of
the orbital angular velocity $m\Omega$.  Lines are as indicated in
Fig.~\protect\ref{fig:Eb_v_J}.}
\label{fig:Eb_v_Omg}
\end{figure}

Figure~\ref{fig:Eb_v_J} compares the numerical results for
binding energy versus orbital angular momentum for the
sequence of quasi-circular orbits against equation (\ref{eqn:pn_EvJ})
with $\eta = 1/4$.
The numerical data is displayed as a bold
solid line with cross marks denoting actual data points.  The
long dashed line corresponds to the Newtonian result, the short
dashed line to the Newtonian result together with the
(post)${}^1$-Newtonian corrections, and the dotted line to the full
(post)${}^2$-Newtonian result.  Notice that for large $J/\mu{m}$
(large separation), the post-Newtonian expansion appears to be
converging quite well toward the numerical result.
Figures~\ref{fig:J_v_Omg} and \ref{fig:Eb_v_Omg} are analogous plots
for the orbital angular momentum versus orbital angular frequency
and binding energy versus orbital angular frequency, respectively.
Again, in the limit of large separation (small $m\Omega$), the
post-Newtonian expansions appear to converge well with the numerical
results.  This agreement at large separations, together with the
general agreement in the shapes of the curves when the separation
decreases, confirm that the basic premises adopted in the definition
of the effective-potential method are sound.  For small separations,
we know that the approximations needed to define the effective
potential as outlined in Sec.~\ref{sec:effect_pot} must be
questioned.  Unfortunately, the post-Newtonian approximation also
breaks down as the separation of the holes becomes small and, it is
impossible to gauge the validity of the approximations based on
the post-Newtonian results.

\section{Discussion}
\label{sec:discussion}
The most important use of the results obtained in this paper is
to narrow the range of initial-data parameters which must be
considered in setting up an actual numerical simulation of the
inspiral and collision of two black holes.  Assuming one is
interested in evolving initial data that represents something
similar to a quasi-circular orbit, then the results presented in
Table~\ref{tab:qso_params} allow us to estimate the minimum length
of time that the system must be evolved, as well as some limits on
the final state of the system.

Beginning with the latter, let us assume that the binary system
coalesces and settles down to a Kerr hole with mass $M_{\rm f}$
and angular momentum $J_{\rm f}$.  From equation (\ref{eqn:m_christ})
we know that the Kerr ratio $J_{\rm f}/M^2_{\rm f}$ is
\begin{equation}
\label{eqn:kerr_ratio}
	{J_{\rm f}\over M^2_{\rm f}} = {{J_{\rm f}/M^2_{\rm ir}}
		\over 1 + {1\over4}\left({J_{\rm f}/
				M^2_{\rm ir}}\right)^2} ,
\end{equation}
where we now let $M_{\rm ir}$ denote the irreducible mass of the
final Kerr hole.  Assuming only that the individual holes do not
rotate, we know that the irreducible mass is bounded by
\begin{equation}
\label{eqn:m_ir}
	M_{\rm ir} \ge \sqrt{M_1^2 + M_2^2} = m\sqrt{1 - 2\eta},
\end{equation}
where $M_1$, $M_2$, etc.\ are defined as before.  Since we know that
$J_{\rm f} \le J$, we find that
\begin{equation}
\label{eqn:Jom_ir2}
	{J_{\rm f}\over M^2_{\rm ir}} \le {\eta\over1 - 2\eta}
		{J\over\mu{m}}.
\end{equation}
Examining equations (\ref{eqn:kerr_ratio}) and (\ref{eqn:Jom_ir2}) we
find that, so long as $J_{\rm f}/M^2_{\rm ir} < 2$, the Kerr
ratio $J_{\rm f}/M^2_{\rm f}$ is {\em guaranteed} to be less than
unity.  For equal-sized holes, this implies that the Kerr ratio is
satisfied as long as $J/\mu{m} < 4$.  If we evolve initial data
representing the innermost stable quasi-circular orbit, we find that
$J_{\rm f}/M^2_{\rm f} \le 0.958$.  If we make the severe assumptions
that all of the binding energy ``released'' in the coalescence is
recaptured by the resulting black hole and that half of the angular
momentum is, nevertheless, radiated away during the final plunge
and coalescence, we find that $J_{\rm f}/M^2_{\rm f} \sim 0.4$.
We see then that a black hole resulting from the inspiral and
final plunge of two nonrotating black holes must certainly be
considered to be rapidly rotating, but it will not violate the
Kerr limit.

Now consider the minimal requirements of a numerical evolution of
binary coalescence in the case of two equal-sized, nonrotating
black holes.  If the simulation is to remotely resemble
the final plunge of two holes following a secular inspiral,
then our best guess at initial data is that
for the innermost stable quasi-circular orbit.  From the estimate
of the orbital angular velocity in Table~\ref{tab:qso_params}, we
find that the orbital period is $\tau \sim 37m$.  It is reasonable
to assume that the final plunge will occur on a time scale comparable
to that of the innermost orbit.  If this initial data leads to an
evolution that begins at the most one orbit before the beginning of
the final plunge and we add to the evolution time a period sufficient
to watch some of the ring-down, then we find that the numerical
simulation must be capable of evolving for 90--130$m$.

\acknowledgments
I gratefully acknowledge many useful discussions with Andrew
Abrahams, Stuart Shapiro, and Saul Teukolsky.  Special thanks
go to Curt Cutler and Lawrence Kidder for helpful discussions
and for contributions to the post-Newtonian comparison.
This work was supported in part by National Science Foundation Grant
No. PHY~90-07834.  Computations were performed at the Cornell
Center for Theory and Simulation in Science and Engineering,
which is supported in part by the National Science Foundation,
IBM Corporation, New York State, and the Cornell Research Institute.

\appendix
\section*{The momenta of individual holes}
A rigorous quasi-local definition of momentum requires the
presence of a Killing vector field $\xi^i_{(k)}$.  The magnitude
of the momentum associated with $\xi^i_{(k)}$ within a given surface
contained in a spatial slice is denoted $\Pi_{(k)}$, where
\begin{equation}
\label{eqn:gen_momentum}
	\Pi_{(k)} \equiv \Pi_i \xi^i_{(k)} = {1\over8\pi}\oint{
		\left(K^j_i - \delta^j_iK\right)\xi^i_{(k)}d^2S_j}.
\end{equation}
Here, $K^i_j$ is the extrinsic curvature of the slice and $K$ is
its trace.  A general spatial slice contains no Killing vectors
and only the total momenta of an asymptotically flat spatial
slice can be defined.  This definition requires integrating
equation (\ref{eqn:gen_momentum}) at spatial infinity using
{\em asymptotic} Killing vectors $\bar{\xi}^i_{(k)}$, which are
Killing vectors of the flat metric to which the physical metric
is asymptotic.  Note, however, that the {\em angular} momentum
will only be gauge-invariant if $\bar{\xi}^i_{(k)}$ is an exact
symmetry of the physical metric (cf York\cite{york80}).

Now consider the momenta of gravitational initial data constructed
within the conformal-imaging approach.  Following this approach, we
conformally decompose the physical metric of a spatial slice as
$\gamma_{ij} = \psi^4\bar{\gamma}_{ij}$, where $\bar{\gamma}_{ij}$ is
the {\em flat} conformal background metric and $\psi$ is the conformal
factor.  With the trace-free conformal background extrinsic curvature
defined by $\bar{A}^i_j = \psi^6(K^i_j - \frac{1}{3}\delta^i_jK)$ and
with $K = 0$, we can rewrite equation (\ref{eqn:gen_momentum}) as
\begin{equation}
\label{eqn:conf_mom}
	\Pi_{(k)} = {1\over8\pi}\oint_\infty{
		\bar{A}^j_i \bar{\xi}^i_{(k)}d^2\bar{S}_j}.
\end{equation}
This form of the equation has the advantage that it does not involve
the conformal factor $\psi$, so we can compute the momentum without
having a solution of the Hamiltonian constraint.  Also, I emphasize
that we have {\em not} used the fact that $\psi\to1$ at spatial
infinity in deriving equation (\ref{eqn:conf_mom}), so that if
$\bar{\xi}^i_{(k)}$ represents an exact symmetry of the {\em physical}
metric, then equation (\ref{eqn:conf_mom}) can be evaluated (for
that Killing vector) by integrating over any two-surface containing
the support of the gravitational field.

The concept of the momenta (linear and angular) of an {\em individual}
hole in the presence of other holes cannot be rigorously defined in
general relativity.  However, a reasonable quasi-local definition for
the momenta of an individual  black hole is equation
(\ref{eqn:conf_mom}) integrated over a two-surface exterior to that
hole.  For evaluating the components of the hole's linear momentum,
we use the three translational Killing vectors of flat Euclidean
3-space, and for the angular momentum, we use the three rotational
Killing vectors with the origin of rotation chosen to be the center
of the hole.

Following the conformal-imaging approach, the background extrinsic
curvature of a spatial slice is constructed as the linear sum of
``single-hole'' extrinsic curvature solutions plus image terms that
maintain an isometry condition (cf Cook \cite{cook91}).  The background
extrinsic curvature for a single black hole (including self-image
terms) is parameterized directly in terms of the physical momenta
measured at infinity.  Evaluation of the black hole's momenta via
the quasi-local definition is independent of the radius of the
surface on which the integral is evaluated and always yields the
correct physical result.
Now consider evaluating the quasi-local momentum integral over a
two-surface that does not contain the black hole.  Using Gauss' law,
we can rewrite the integral as
\begin{equation}
\label{eqn:vol_mom}
	\Pi_{(k)} = {1\over8\pi}\int{\bar\nabla_j\left[
		\bar\xi^i_{(k)}\bar{A}^j_i\right]d^3\bar{V}} = 0,
\end{equation}
since $\bar\xi^i_{(k)}$ is a Killing vector of $\bar\gamma_{ij}$ and
$\bar{A}^j_i$ satisfies the {\em vacuum} momentum constraint equation
in this volume.  Constructing a multi-hole extrinsic curvature
from single-hole extrinsic curvature solutions, including self-image
terms but not including general image terms, we see that equation
(\ref{eqn:vol_mom}) implies that the contributions to the extrinsic
curvature from additional holes do not effect the quasi-local
momenta of a given hole.

What remains is to examine the contribution of general image terms
to the quasi-local momenta of a hole.  It seems reasonable that
these terms should have no contribution, however, I have so far
been unable to prove this analytically.  Fortunately, it is
straightforward to show numerically that general image terms make
no contribution to the quasi-local momenta of either hole in a
general binary configuration.  More specifically, we can construct
a general solution for the background extrinsic curvature including
any number of image terms.  Computing the quasi-local momenta for
either of the holes, we can use Richardson extrapolation to show
that, up to the numerical precision of the computer, the results
are {\em identical} to those obtained in the absence of any general
image terms.  This is independent of the sizes and separations of
the holes and of the number of image terms included in the extrinsic
curvature implying that each image term {\em independently}
contributes nothing to any of the surface integrals.  I suspect
that this result holds for any number of holes, however, this has
not been verified.

We see then that within the limitations of defining the momenta
of an individual hole, the momenta used to parameterize a single
hole {\it are} the momenta of individual holes within a multi-hole
configuration.  That is, linear or angular momenta on one hole
do not induce any amount of linear or angular momentum on any other
hole in the system.  In particular, the orbital angular momentum of
a system of holes is well defined and does not contain an induced
spin on any of the holes due to the linear momenta of the holes.


\begin{references}
\bibitem{cook93}
G.~B. Cook {\it et~al.}, Phys.\ Rev.\ D {\bf 47},  1471  (1993).
\bibitem{york79}
J.~W. York, Jr.,  in {\em Sources of Gravitational Radiation},
edited by L.~L. Smarr (Cambridge University Press, Cambridge,
England, 1979), pp.\ 83--126.
\bibitem{bowen79}
J.~M. Bowen, Gen.\ Rel.\ Grav. {\bf 11},  227  (1979).
\bibitem{bowenyork80}
J.~M. Bowen and J.~W. York, Jr., Phys.\ Rev.\ D {\bf 21},  2047  (1980).
\bibitem{bowen82}
J.~M. Bowen, Gen.\ Rel.\ Grav. {\bf 14},  1183  (1982).
\bibitem{ksy83}
A.~D. Kulkarni, L.~C. Shepley, and J.~W. York, Jr., Phys.\ Lett.
{\bf 96A}, 228  (1983).
\bibitem{cook91}
G.~B. Cook, Phys.\ Rev.\ D {\bf 44},  2983  (1991).
\bibitem{ADM}
R. Arnowitt, S. Deser, and C.~W. Misner,  in {\em Gravitation: An
Introduction to Current Research}, edited by L. Witten (Wiley, New
York, 1962), pp.\ 227--265.
\bibitem{detweiler92}
J.~K. Blackburn and S. Detweiler, Phys.\ Rev.\ D {\bf 46},  2318
(1992).
\bibitem{kidder92}
L.~E. Kidder, C.~M. Will, and A.~G. Wiseman, Class.\ Quantum\ Grav.
{\bf 9}, L125  (1992).
\bibitem{Frontiers:York}
J.~W. York, Jr.,  in {\em Frontiers in Numerical Relativity}, edited
by C.~R.  Evans, L.~S. Finn, and D.~W. Hobill (Cambridge University
Press, Cambridge, England, 1989), pp.\ 89--109.
\bibitem{christ70}
D. Christodoulou, Phys.\ Rev.\ Lett. {\bf 25},  1596  (1970).
\bibitem{cook90}
G.~B. Cook and J.~W. York, Jr., Phys.\ Rev.\ D {\bf 41},  1077  (1990).
\bibitem{cookabrahams92}
G.~B. Cook and A.~M. Abrahams, Phys.\ Rev.\ D {\bf 46},  702  (1992).
\bibitem{anninos_etal94}
P. Anninos {\it et~al.}, Phys.\ Rev.\ Lett.  submitted  (1994).
\bibitem{kidder93}
L.~E. Kidder, C.~M. Will, and A.~G. Wiseman, Phys.\ Rev.\ D {\bf 47},
3281 (1993).
\bibitem{york80}
J.~W. York, Jr.,  in {\em Essays in General Relativity}, edited by
F.~J. Tipler (Academic Press, New York, 1980), pp.\ 39--58.
\end{references}

\end{document}